%% file: main.tex
\documentclass[10pt,
twocolumn,letterpaper]{article}

\usepackage{cvpr} 
\usepackage{times}
\usepackage{balance}
\usepackage{epsfig}
\usepackage{graphicx}
\usepackage{amsmath}
\usepackage{amssymb}
\usepackage{subcaption}
\usepackage{color}
\usepackage{comment}
\usepackage{threeparttable}
\usepackage{soul}
\usepackage{diagbox}
\usepackage{multirow}
\usepackage{makecell}
\usepackage{booktabs}
\usepackage{colortbl}
\usepackage{siunitx}
\usepackage{algorithm}
\usepackage[table]{xcolor}
\usepackage{algpseudocode}
\usepackage{xr}

\definecolor{lightgray}{gray}{0.9}
\definecolor{lightblue}{RGB}{224, 235, 255}

\makeatletter
\def\UrlAlphabet{%
      \do\a\do\b\do\c\do\d\do\e\do\f\do\g\do\h\do\i\do\j%
      \do\k\do\l\do\m\do\n\do\o\do\p\do\q\do\r\do\s\do\t%
      \do\u\do\v\do\w\do\x\do\y\do\z\do\A\do\B\do\C\do\D%
      \do\E\do\F\do\G\do\H\do\I\do\J\do\K\do\L\do\M\do\N%
      \do\O\do\P\do\Q\do\R\do\S\do\T\do\U\do\V\do\W\do\X%
      \do\Y\do\Z}
\def\UrlDigits{\do\1\do\2\do\3\do\4\do\5\do\6\do\7\do\8\do\9\do\0}
\g@addto@macro{\UrlBreaks}{\UrlOrds}
\g@addto@macro{\UrlBreaks}{\UrlAlphabet}
\g@addto@macro{\UrlBreaks}{\UrlDigits}
\makeatother

\newcommand{\includegraphicsorplaceholder}[2][]{%
  \IfFileExists{#2}{%
    \includegraphics[#1]{#2}%
  }{%
    \setlength{\fboxsep}{0pt}%
    \fbox{%
      \begin{minipage}[c][100pt][c]{100pt}
        \centering
        \textit{Image not found:}\\
        \texttt{#2}
      \end{minipage}%
    }%
  }%
}

\usepackage[pagebackref=true,breaklinks=true,letterpaper=true,colorlinks,bookmarks=false]{hyperref}

\begin{document}

\title{WaveVerify: A Novel Audio Watermarking Framework for Media Authentication and Combatting Deepfakes}

\author{Aditya Pujari and Ajita Rattani\\
University of North Texas\\
Denton, Texas, USA\\
{\tt\small adityapujari@my.unt.edu; ajita.rattani@unt.edu}
}

\maketitle
\thispagestyle{empty}

\begin{abstract}
With the rise of voice synthesis technology threatening digital media integrity, audio watermarking has become a crucial defense for content authentication. However, current solutions often lack robustness to effects like high-pass filtering and temporal modifications and suffer from poor watermark localization. We introduce \textbf{WaveVerify}, a watermarking system that addresses these challenges using a Feature-wise Linear Modulation (FiLM)-based generator for resilient multiband watermark embedding and a Mixture-of-Experts detector for accurate extraction and localization. Our unified training framework enhances robustness by applying multiple distortions per backpropagation step via a dynamic effect scheduler. Evaluations across multiple datasets show WaveVerify outperforms SOTA models like AudioSeal and WavMark, achieving zero Bit Error Rate (BER) under common distortions and MIoU scores of 0.98+ under severe temporal modifications. The parallel FiLM-based generator also reduces training time by $\sim$80\% compared to sequential embedding approaches. Code and pretrained models are available at: \url{https://github.com/vcbsl/WaveVerify}.

\end{abstract}

\section{Introduction}

The rapid advancement of voice generation technologies has enabled the synthesis of speech that is perceptually indistinguishable from genuine human voices~\cite{valle, speartts}. While these innovations facilitate beneficial applications such as personalized text-to-speech systems~\cite{yan1} and voice preservation~\cite{voicepreservation}, they have also introduced significant risks, including deepfake impersonation scams~\cite{yan2} and synthetic media-driven disinformation campaigns~\cite{baigong}. Recent reports indicate that in $2024$, deepfake fraud attempts surged by over 1,300\% compared to 2023~\cite{pindrop2025}, underscoring the urgent need for robust audio content authentication. The financial sector has been particularly impacted, with a loss of over \$10 million to voice scams~\cite{financialbrand2025} and individual victims reporting losses exceeding \$6,000 from AI-generated deepfake calls~\cite{businesswire2025}. In response, regulators and governments worldwide are enacting measures to improve AI content transparency and traceability, emphasizing the development of forensic tools and watermarking techniques as essential strategies to uphold media integrity.

\setcounter{figure}{1}

\begin{figure*}[t]
  \centering
  \begin{subfigure}[t]{0.70\textwidth}
    \centering
    \includegraphics[width=\linewidth]{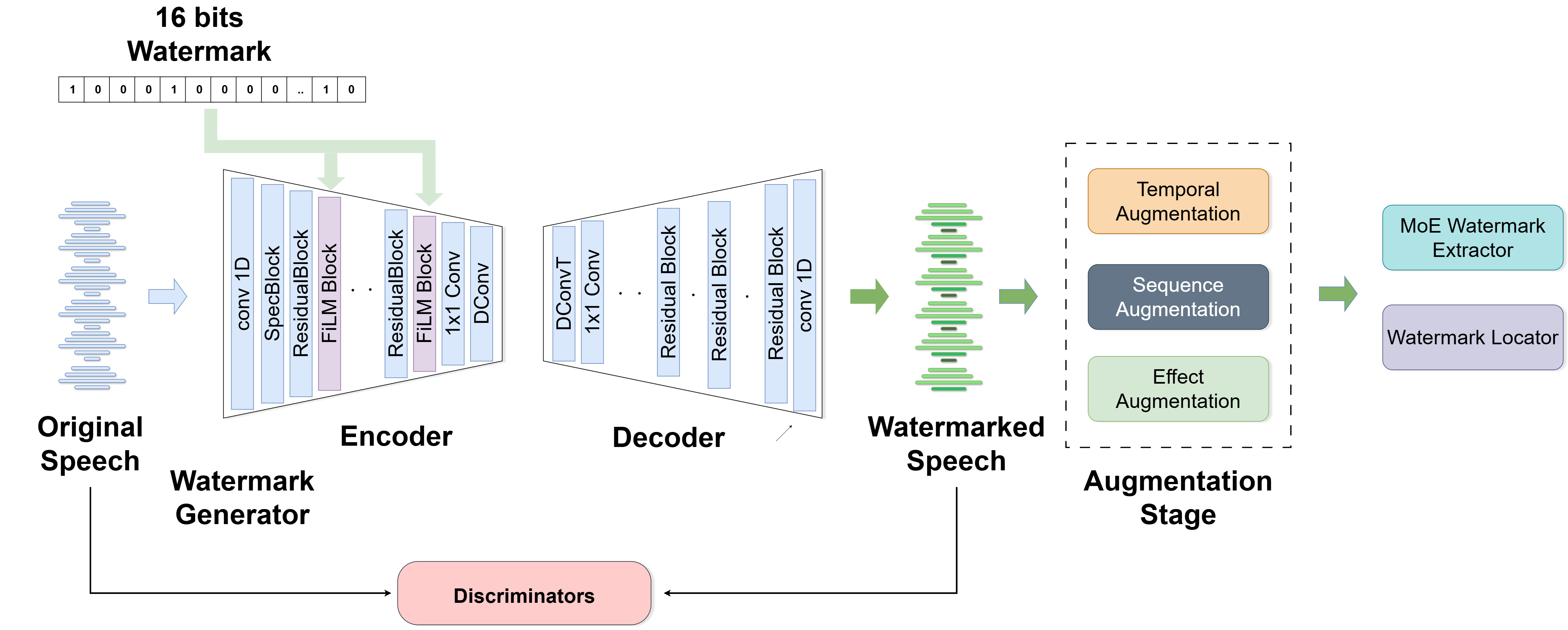}
    \caption{End-to-end Training Pipeline.}
    \label{fig:end-to-end-architecture-a}
  \end{subfigure}
  \hfill
  \begin{subfigure}[t]{0.25\textwidth}
    \centering
    \includegraphics[width=\linewidth]{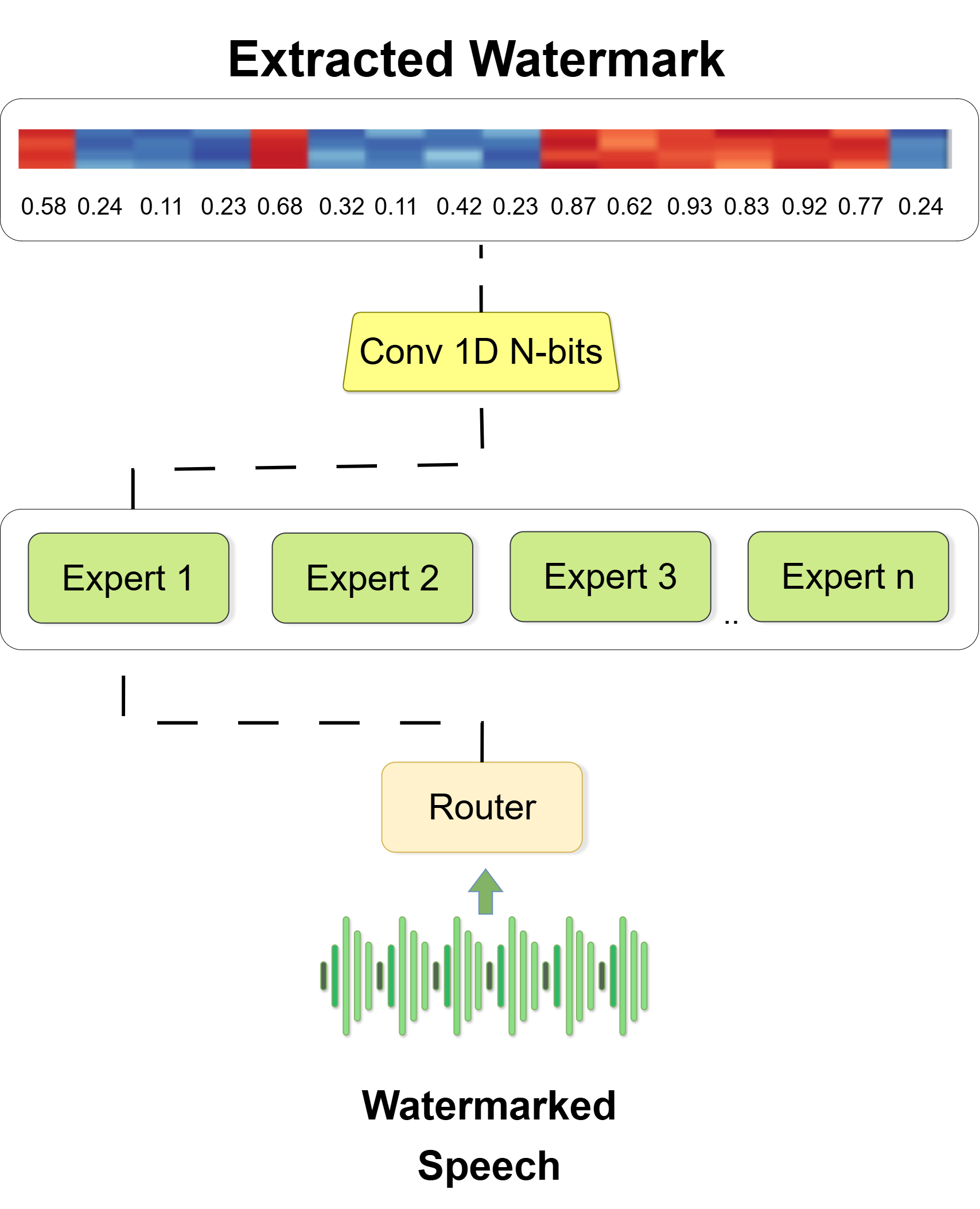}
    \caption{MoE Watermark Extractor.}
    \label{fig:end-to-end-architecture-b}
  \end{subfigure}

  \caption{%
    (a) The end-to-end training pipeline of WaveVerify, where a FiLM-based generator embeds bits into speech, followed by temporal/sequence/effect augmentations, and extraction via a locator and MoE detector. 
    (b) MoE watermark detector architecture: Input audio is processed by a HILCodec encoder, then routed through multiple expert networks via a learned gating mechanism to produce the final extracted watermark bits.
  }
  \label{fig:end-to-end-architecture}
\end{figure*}

A predominant forensic approach for detecting synthesized audio involves training binary classifiers to distinguish between natural and synthetic speech, as demonstrated in~\cite{xiao2024xlsr, zhang2024audio}. This method, commonly referred to as \textit{passive detection}, offers a relatively straightforward mitigation strategy. However, its effectiveness is increasingly challenged as generative models continue to improve, narrowing the perceptual and statistical gap between authentic and synthetic audio.

Audio watermarking has thus emerged as a promising alternative for asserting ownership and authenticity of audio media. By embedding imperceptible yet robust signals within audio content, watermarking enables reliable detection and verification~\cite{audioseal,wavmark}. Deep learning-based watermarking methods have addressed many limitations of traditional techniques (e.g., spread spectrum~\cite{cox1997secure}, patchwork~\cite{2003patchwork}, and echo hiding~\cite{echo}), improving imperceptibility, retrieval accuracy (measured by Bit Error Rate, BER), and resilience to audio effects. These methods typically employ a generator for watermark embedding, an effect simulator for robustness, and a detector for retrieval~\cite{zhu2018hidden, Deorustc, silentcipher}. However, most rely on audio effect simulators with fixed distortion parameters, which can limit effectiveness against unforeseen real-world attacks (audio modifications/ manipulations).

Recent innovations such as WavMark~\cite{wavmark} and AudioSeal~\cite{audioseal} have improved these foundations. WavMark introduced synchronization patterns for watermark localization, while AudioSeal achieved faster detection speeds and enhanced sample-wise watermark localization. Despite these advances, current watermarking techniques \emph{remain vulnerable} to temporal manipulations (e.g., reversal and speed changes), structural modifications (e.g., trimming and segment shuffling), and obtain limited localization precision, which undermines their effectiveness in adversarial or real-time scenarios.

WaveVerify addresses these aforementioned limitations by rethinking watermark embedding and detection. It uses a Feature-wise Linear Modulation (FiLM)-based generator to adaptively spread watermark features across multiple frequency bands and temporal scales, enhancing robustness against spectral and temporal attacks. For detection, a Mixture-of-Experts (MoE) architecture routes extraction through specialized sub-networks resilient to distortions. Its training pipeline includes aggressive, adaptive augmentations, enabling reliable detection even under severe audio alterations.

The key \textbf{innovations} of WaveVerify are listed as follows:

\begin{itemize}
    \item \textbf{A robust FiLM-based generator architecture} that adaptively and uniformly embeds and distributes watermark features across multiple frequency bands and temporal scales to ensure robustness to audio effects and temporal manipulations.
    \item \textbf{A Mixture-of-Experts detector} for resilient watermark extraction by dynamically routing through specialized sub-networks tailored to diverse distortions and manipulations, consequently enhancing robustness to a wide range of unseen audio manipulations.
    \item \textbf{A dynamic augmentation and effect scheduling strategy} that accelerates training and ensures robustness to a wide range of real-world audio effects or distortions.
    \item \textbf{Comprehensive empirical validation} demonstrating state-of-the-art performance in both watermark bit recovery (BER) and localization (MIoU), with significant improvements and robustness against attacks over prior work.
\end{itemize}

\section{Related Work}

Audio watermarking has evolved significantly, beginning with traditional signal processing techniques that laid the foundation for imperceptible and robust embedding. Early approaches, such as spread spectrum watermarking, distributed signals across the frequency domain to resist interference~\cite{cox1997secure}, while echo hiding techniques embedded watermarks by introducing imperceptible echoes with specific delay parameters~\cite{boney1996digital}. Phase-based methods further advanced robustness by modifying phase components of audio signals~\cite{arnold2003phase}. However, these methods struggled against modern audio transformations like compression and filtering, as highlighted by Cox et al.~\cite{cox2002digital} and Bassia et al.~\cite{bassia2001robust}, underscoring the need for more adaptive solutions.

The advent of deep learning revolutionized audio watermarking by enabling end-to-end optimization of watermark embedding and extraction processes. HiDDeN introduced a pioneering framework for a trainable watermarking solution, leveraging neural networks to enhance robustness against diverse distortions~\cite{zhu2018hidden}. More recent deep learning-based methods have further improved adaptability and performance by jointly training generator and detector networks, often incorporating adversarial objectives and distortion simulation during training~\cite{SoK2024,EMNLP2024}. For instance, IDEAW employs invertible dual-embedding and attack-aware training to improve both capacity and localization, addressing key challenges in neural watermarking~\cite{EMNLP2024}.

Recent techniques like WavMark~\cite{wavmark} and AudioSeal~\cite{audioseal} have further improved performance through synchronization patterns and joint optimization of generator-detector networks, respectively. WavMark leverages invertible neural networks for reciprocal encoding and decoding, achieving high capacity and imperceptibility, as well as automatic watermark localization. AudioSeal achieves superior temporal precision and detection speeds but faces challenges with specific attack vectors like high-pass filtering or temporal modifications. Benchmarking studies such as AudioMarkBench confirm that while these methods are robust to many black-box forgery attacks, they remain vulnerable to adaptive watermark-removal strategies, especially those that exploit detector APIs or operate in the spectrogram domain~\cite{AudioMarkBench2024}.

Despite these advancements, existing watermarking techniques remain vulnerable, particularly to temporal and combined attacks that involve multiple types of distortions (e.g., speed changes combined with spectral filtering, or compression followed by noise addition)~\cite{SoK2024,AudioMarkBench2024}. Moreover, recent work has highlighted the limitations of current schemes in the face of generative AI-driven attacks, which can remove watermarks while preserving audio quality, raising concerns about the long-term viability of watermarking for intellectual property protection~\cite{SoK2024}. While solutions like SilentCipher~\cite{silentcipher} have explored information-theoretic capacity optimization and adversarial training to counter unseen distortions, their practical application is limited by constraints such as a lack of explicit localization capabilities. These limitations motivate the development of WaveVerify, which addresses these gaps through a novel FiLM-based generator architecture and Mixture-of-Experts detector design.

\section{Proposed Method}

\subsection{WaveVerify Framework Overview}

Figure~\ref{fig:end-to-end-architecture} shows the overview of the framework of WaveVerify. The proposed WaveVerify framework introduces a novel end-to-end approach for robust audio watermarking, integrating three synergistic components within a unified Audio Watermarking architecture as illustrated in Figure~\ref{fig:end-to-end-architecture}(a). At its core, the framework consists of a Generator, a Locator, and a Detector, each optimized for specific tasks in the watermarking pipeline.

The Generator trained adversarially alongside a Discriminator, employs an encoder-decoder architecture with Feature-wise Linear Modulation (FiLM)~\cite{film} for watermark embedding. This design choice is crucial for computational efficiency, as conventional bottleneck embedding approaches require significantly more training time due to the sequential nature of processing through the entire network. By leveraging FiLM for adaptive modulation at multiple hierarchical levels, our approach distributes the watermark embedding process across the network architecture, enabling more efficient parallel computation and reducing overall training time by approximately 80\% compared to our implemented bottleneck-based alternatives (embedding watermark at the bottleneck layers) while maintaining robust watermark integration. Direct training time comparisons with existing methods, such as AudioSeal, are not feasible due to variations in hardware configurations and the lack of publicly available or executable implementation code.

The framework's Locator enables precise identification of watermark presence or absence for each individual audio sample (i.e., sample-level identification) of watermarked regions while maintaining minimal computational overhead. Complementing these components, the Detector utilizes a sophisticated architecture (approximately $4.5$M parameters) featuring a Mixture-of-Experts (MoE) module with multiple specialized expert networks (referred to as sub-networks), significantly enhancing the model's capacity to diverse audio attacks (detailed in Figure~\ref{fig:end-to-end-architecture}(b)).

To further improve robustness against audio edits and reduce training time, the framework incorporates temporal augmentations and audio effects, using a dynamic effect scheduler, during the end-to-end training stage. Next, we discuss these individual components of the WaveVerify framework as follows:

\subsection{Watermark Embedding and Modulation}

Previous watermarking approaches like AudioSeal~\cite{audioseal} use "direct linear extension" to embed watermarks by simply repeating fixed message-derived vectors across the temporal dimensions of intermediate audio features. This method lacks resilience to temporal modifications in audio, resulting in suboptimal and uneven distribution of watermark information across different time segments. We embed watermarks across the entire audio signal without using voice activity detection. Addressing these limitations, we introduce a sophisticated hierarchical modulation architecture that operates at multiple temporal scales as follows:

\begin{itemize}

\item A message processing module employs a multi-layer perceptron (MLP) to transform the $n$-bit watermark sequence into adaptive modulation parameters (scale $\gamma$ and shift $\beta$ vectors), enabling dynamic control over the watermark embedding process
\item The framework implements multi-scale feature modulation throughout the encoder's hierarchical layers, utilizing frequency-specific FiLM (Feature-wise Linear Modulation)~\cite{film} layers to apply adaptive modulation across multiple frequency bands, enhancing the robustness of the watermarking to temporal edits and audio effects such as high pass filtering. This multi-band distribution prevents the concentration of watermark information in specific frequency regions, thereby increasing resilience against filtering attacks that target particular frequency ranges.
\end{itemize}

FiLM modulates features directly ($F' = \gamma \odot F + \beta$) without normalization, preserving signal statistics crucial for watermarking, unlike AdaIN (Adaptive Instance Normalization) which first normalizes features to zero mean and unit variance before applying affine transformations, potentially removing important signal characteristics needed for robust watermark embedding.
WaveVerify leverages the HILCodec architecture~\cite{hilcodec}, a state-of-the-art neural audio codec recognized for its high-fidelity output and computational efficiency across various bitrates. The generator adopts HILCodec's variance-constrained design with time-domain fully convolutional layers, incorporating spectrogram blocks and carefully designed residual components to extract rich feature representations while maintaining minimal processing overhead. This architecture's multi-resolution capability and L2-normalization techniques naturally complement WaveVerify's FiLM-based modulation approach, allowing watermark information to be adaptively distributed across multiple frequency bands and temporal scales. For additional technical details regarding the FiLM-based embedding approach, refer to the supplementary material (Appendix A). By inheriting HILCodec's depthwise-separable convolutions and optimized encoder-decoder structure, WaveVerify achieves exceptional robustness against spectral filtering attacks while maintaining imperceptible watermarking with significantly reduced computational requirements compared to previous approaches.

\subsection{Augmentation Stage}
To enhance watermark robustness against a wide range of potential modifications, we employ a two-level augmentation strategy during the training stage, explained as follows:

\subsubsection{Temporal Augmentations}
\label{sec:temporal_structural_augmentation_combined}
At this stage, we add two kinds of temporal augmentations, namely, segment-level transformations targeting localized regions and sequence-level transformations altering the entire temporal structure.

For segment-level temporal augmentations, the framework operates on fixed-duration audio segments (0.1s) and modifies 20\% of randomly selected segments, applying with equal probability one of three transformations: replacing watermarked segments with non-watermarked counterparts, setting segments to silence, or replacing segments with audio from a different source. Complementing these, we implement sequence-level augmentations by randomly applying one transformation to the entire audio signal while preserving watermark content: reversing the temporal order, rotating by a random offset, or shuffling fixed-length segments (e.g., 0.5s). By forcing the model to identify watermarks across varied sequential patterns, it learns intrinsic features rather than positional cues, enabling robust sample-level detection even under significant reordering.
\vspace{-0.5em}
\subsubsection{Audio Effect Augmentation}
The second class of augmentation ensures robustness against audio editing. To simulate real-world modifications, our augmentation pipeline includes diverse audio effects: 
highpass filter, lowpass filter, bandpass filter, resample, speed modification, and random noise  (details in Appendix B.1). 

The parameters of those augmentations are fixed to aggressive values to enforce maximal robustness, and the probability of sampling a given augmentation is proportional to the inverse of its evaluation detection accuracy. 
Instead of employing fixed augmentation parameters and static selection probabilities, we introduce a Dynamic Effect Scheduler, inspired by curriculum learning principles~\cite{Bengio2009Curriculum} but specifically adapted for adversarial watermarking. This scheduler intelligently manages the audio effect augmentation pipeline during training. It adaptively adjusts both the selection probability and the specific parameters (e.g., filter cutoff frequencies, noise levels) for each audio effect. This adaptation is driven by real-time performance metrics, primarily the Bit Error Rate (BER) and Mean Intersection over Union (MIoU), calculated on the augmented samples. By prioritizing effects and parameter settings that currently pose a greater challenge to the model (indicated by a higher BER or lower MIoU), the scheduler ensures that the model progressively develops robustness against the most difficult transformations. The precise mechanism, including the use of exponential moving averages for metric smoothing and the formula for updating probabilities and parameter distributions, is detailed in Appendix B.2.

Importantly, all augmentations are applied on-the-fly during training to each batch of audio samples, rather than pre-generating augmented datasets. This dynamic approach ensures diverse training conditions while maintaining memory efficiency, though it introduces approximately 15\% computational overhead compared to training without augmentations due to the real-time audio processing operations.

\subsection{Dual-Network Architecture for Watermark Detection and Localization}

The architecture employs two specialized neural networks with shared design principles but distinct functional objectives: (1) a \textbf{Detector Network} for robust message recovery through multi-scale feature analysis, which includes a gating router component to predict and select the most suitable expert networks (using MOE) for processing a given input, and (2) a \textbf{Locator Network} for precise temporal identification of watermarked regions via high-resolution processing. This bifurcated approach addresses fundamental technical constraints in audio watermarking rooted in well-documented signal processing principles\cite{cox2002digital}, namely, (a) the feature complexity vs. resolution trade-off, as detection requires rich feature representations (128+ channels) while localization demands lightweight architectures (64 channels) to preserve temporal resolution, and (b) computational efficiency considerations, as a unified network would require maintaining high channel dimensions throughout to support detection while also preserving full temporal resolution for localization. For additional theoretical justification and empirical results supporting this dual-network design, please refer to the supplementary material, Appendix C.

\noindent \textbf{Architectural Details.} The Detector implements a hierarchical encoder-decoder structure using a modified SEANet (Speech Enhancement Audio Network)\cite{seanet} encoder with (a) four convolutional blocks with kernel size=7, stride=\{2,2,2,2\}, and channel dimensions=\{128,256,384,512\}, (b) GroupNorm normalization and Mish activations ($f(x) = x \cdot \tanh(\ln(1+e^x))$) for improved gradient flow and training stability, and (c) a decoder with four experts ($E=4$).

The decoder employs a Mixture of Experts (MoE) approach. Each expert $f_i$ consists of three depthwise-separable convolutional layers (kernel size=3, stride=1) followed by temporal unpooling. The final output combines the predictions from all experts through a weighted sum using learned gating weights, applied via the Hadamard product ($\odot$):

\begin{equation}
\text{Output} = \sum_{i=1}^4 \sigma(W_g^{(i)}z) \odot f_i(z)
\label{eq:moe_output}
\end{equation}

where $z$ is the encoded representation from the encoder, $f_i(z)$ is the output of the $i$-th expert network, $W_g^{(i)}$ are the learned gating weights for expert $i$, and $\sigma$ represents the sigmoid activation function. At inference, the gating network output, $\sigma(W_g^{(i)}z)$, effectively acts as a selection probability or weight, determining the contribution of each expert $f_i$ to the final output based on the input representation $z$.

The Locator employs a streamlined SEANet ~\cite{seanet} variant optimized for temporal precision with (a) three convolutional blocks with kernel size=7, stride=\{2,2,2\}, and channel dimensions=\{64,64,64\}, (b) transposed convolutions with stride=\{8,4,2\} for resolution recovery, and (c) depthwise-separable convolutions reduce parameters by 7$\times$ versus standard convolutions.

\textbf{Parameter Efficiency Techniques.} The locator network employs several techniques to achieve significant parameter efficiency compared to the Detector expert architecture, resulting in a dramatic reduction from 4.5M parameters (Detector Expert) to just 0.13M parameters (Locator), a 97\% reduction. This efficiency is crucial for maintaining high temporal resolution for localization. 
Using $64$ vs. $128$ channels results in 4$\times$ fewer parameters while maintaining sufficient feature expressivity. A lightweight multi-scale feature fusion mechanism that combines features from different resolution levels (stride-2, stride-4, and stride-8) using 1$\times$1 convolutions before upsampling, improving accuracy while adding minimal parameters. Specific encoder layers share weights across similar processing stages, reducing unique parameters by 15\% without significant performance impact.

Further, through symmetric padding and stride management, the locator network maintains strict input-output temporal alignment:
\begin{equation}
T_{out} = \left\lfloor \frac{T_{in} + 2P - K}{S} \right\rfloor + 1 \quad \text{(per block)}
\end{equation}

where $P$ represents the padding, $K$ is the kernel size, and $S$ is the stride. By carefully selecting values for each layer (P=3, K=7, S=2 for encoding layers; P=1, K=4, S=2 for decoding layers), this precise temporal alignment enables detection windows as small as 50ms with positional accuracy of $\pm$2 samples at 16kHz, critical for identifying partial tampering. 

\subsection{Loss Functions}
WaveVerify combines reconstruction, detection, localization, and adversarial losses into a multi-objective function optimized through hierarchical weighting. 

\noindent Reconstruction and Localization Loss. Building on spectral reconstruction losses from neural codecs~\cite{codecwithimprovedrvqgan}, we introduce a weighted scheme prioritizing time-domain fidelity:
\begin{equation}
\resizebox{\linewidth}{!}{$
\mathcal{L}_{\text{rec}} = \lambda_{\text{wave}}\mathcal{L}_{\text{wave}} + \lambda_{\text{spec}}\mathcal{L}_{\text{spec}} + \lambda_{\text{mel}}\mathcal{L}_{\text{mel}}
$}
\end{equation}
where $\mathcal{L}{\text{wave}} = |x-\hat{x}|1$ represents direct waveform matching between original ($x$) and watermarked ($\hat{x}$) audio, $\mathcal{L}{\text{spec}}$ computes multi-scale STFT loss (window sizes 512/2048), and $\mathcal{L}{\text{mel}}$ calculates mel-band (150/80 filter banks) spectral loss. Our weighting scheme emphasizes temporal accuracy over spectral fidelity based on ablation studies.

\noindent Detection Loss. For detection, we jointly optimize two complementary BCE losses, i.e. presence-masked detection loss and temporal localization loss. The presence-masked detection loss focuses learning on watermarked regions using binary mask $m_i \in {0,1}$:
\begin{equation}
\resizebox{\linewidth}{!}{$
\mathcal{L}_{\text{det}} = -\frac{1}{N}\sum_{i=1}^N m_i\left[y_i\log p_i^{\text{(det)}} + (1-y_i)\log(1-p_i^{\text{(det)}})\right]
$}
\end{equation}
where $p_i^{\text{(det)}}$ represents the detector network's predicted probability of watermark bit value at position $i$, and $y_i \in {0,1}$ denotes ground truth watermark bits. This formulation concentrates gradient updates on areas containing actual watermark content.

The temporal localization loss enforces global awareness by predicting watermark presence across all timesteps:

\begin{equation}
\resizebox{\linewidth}{!}{$
\mathcal{L}_{\text{loc}} = -\frac{1}{N}\sum_{i=1}^N\left[m_i\log p_i^{\text{(loc)}} + (1-m_i)\log(1-p_i^{\text{(loc)}})\right]
$}
\end{equation}

Here $p_i^{\text{(loc)}}$ denotes the locator network's prediction of watermark presence at position $i$, while $m_i$ serves as the ground truth mask. Though structurally similar to $\mathcal{L}_{\text{det}}$, this loss optimizes temporal boundary detection rather than bit value accuracy, requiring separate network heads.

\noindent Adversarial Loss. Adversarial training employs multi-scale discriminators with weighted objectives:

\begin{equation}
\resizebox{\linewidth}{!}{$
\mathcal{L}_{\text{adv}} = \lambda_{\text{gen}}\mathbb{E}\left[(1-D(G(x)))^2\right] + \lambda_{\text{feat}}\sum_{l=1}^L|f_l(x)-f_l(G(x))|_1
$}
\end{equation}

where $f_l(x)$ represents feature maps at layer $l$ of discriminator $D$ when processing real audio $x$, and $f_l(G(x))$ corresponds to audio features of the generated audio. The feature matching term stabilizes training by preserving intermediate representations across multiple discriminator layers ($L$).

\noindent Combined Loss. The complete objective integrates these components through task-specific weights:

\begin{equation}
\resizebox{\linewidth}{!}{$
\mathcal{L}_{\text{total}} = \mathcal{L}_{\text{rec}} + \lambda_{\text{det}}\mathbb{E}_\epsilon[\mathcal{L}_{\text{det}}^\epsilon] + \lambda_{\text{loc}}\mathbb{E}_\epsilon[\mathcal{L}_{\text{loc}}^\epsilon] + \mathcal{L}_{\text{adv}}
$}
\end{equation}

The expectation $\mathbb{E}_\epsilon$ computes average over augmentation variants from our dynamic scheduler, implemented through parallel processing of augmented copies per sample. This weighting prioritizes detection robustness (BER) over localization precision (MIoU).
\section{Experiments}

\subsection{Dataset}

To achieve domain generalization in speech processing, we utilize multiple diverse speech datasets~\cite{librispeech, commonvoice, cmuarctic, dipco}, all resampled to 16kHz. We implement stratified batch sampling with controlled distribution mentioned in percentage:

\begin{itemize}
\item LibriSpeech (40\%) \cite{librispeech}: 1,000 hours of read English speech, with the main corpus used for training and a held-out set of 100 unseen speakers reserved for testing.
\item Common Voice (30\%) \cite{commonvoice}: 200 hours spanning 10 languages, with the bulk used for training and 5,000 test clips from 1,000 new speakers held out for testing.
\item CMU ARCTIC (15\%) \cite{cmuarctic}: 20 hours of professional speech, with the majority used for training and testing performed using a leave-two-speakers-out strategy.
\item DiPCo (15\%) \cite{dipco}: 40 hours of conversational speech, with the initial sessions used for training and the last 10 sessions reserved for testing.
\end{itemize}

This multi-dataset approach, combined with strict speaker-disjoint testing within each primary dataset, helps ensure model generalization. Furthermore, we assessed cross-domain performance on entirely unseen datasets, RAVDESS~\cite{ravdess} and ASVspoof 2019~\cite{asvspoof2019}, confirming robust generalization.

\subsection{Experiment Setup}

The experiments were conducted using the PyTorch framework on NVIDIA Quadro RTX 8000 GPUs. All audio samples were preprocessed to 16kHz sampling rate. The training set comprised 500,000 audio segments of 1.0-second duration, while validation utilized 50 distinct segments. For comprehensive evaluation, we employed 1,000 extended audio clips of 10.0 seconds each from RAVDESS and ASVspoof datasets.
 
For training the proposed WaveVerify watermarking model, the optimization protocol employed AdamW optimizer with hyperparameters $\beta_{1}=0.8$, $\beta_{2}=0.99$, and an initial learning rate of $1\times 10^{-4}$, coupled with an exponential decay schedule ($\gamma=0.999996$).
The model training was carried out for $600{,}000$ iterations with a batch size of $32$. Validation was performed at $1{,}000$-iteration intervals.

In addition to evaluating on held-out portions of the primary datasets, we performed cross-domain evaluation using completely unseen test datasets: RAVDESS~\cite{ravdess} and ASVspoof 2019~\cite{asvspoof2019}. These datasets feature different acoustic conditions, recording environments, and speaker demographics than those used during training, allowing for a true assessment of model generalization. This strict speaker-disjoint and cross-domain evaluation ensures that performance metrics reflect genuine generalization rather than memorization of speaker characteristics or dataset-specific artifacts.

\subsection{Evaluation Metrics}
The system's performance is evaluated using two complementary metrics: the standard Bit Error Rate (BER) for message recovery accuracy, and we employ Mean Intersection over Union (MIoU) to assess the precision of watermark localization. To comprehensively assess the perceptual quality of the watermarked audio, we also utilize industry-standard metrics including Perceptual Evaluation of Speech Quality (PESQ)~\cite{pesq} for perceived speech quality, Short-Time Objective Intelligibility (STOI)~\cite{stoi} for intelligibility, Signal-to-Interference Ratio (SISNR) for signal clarity, and ViSQOL Audio Quality Metric (VISQOL)~\cite{visqol} for overall audio quality. This comprehensive evaluation framework ensures both reliable message extraction and accurate temporal identification of watermarked regions, addressing the dual challenges of robust watermark detection and precise localization.

\section{Results}

In order to provide a transparent and comprehensive evaluation, all experiments were repeated in triplicate. Results are reported as mean $\pm$ standard deviation, and paired t-tests confirm that improvements (e.g., in MIoU and BER) are statistically significant (all $p<0.001$).

\subsection{Audio Quality Assessment and Trade-Off Analysis}

Table~\ref{tab:audio_quality} reports various audio quality metrics including PESQ, STOI, ViSQOL, and SISNR, measured on the unseen test set of 10-second clips. WaveVerify achieves perfect speech intelligibility (STOI = 1.00 $\pm$ 0.00) and the highest perceptual quality (ViSQOL = 4.76 $\pm$ 0.07). AudioSeal records the highest PESQ (4.59 $\pm$ 0.06), while WavMark demonstrates superior signal fidelity (SISNR = 36.28 $\pm$ 0.50 dB). These differences highlight design trade-offs between perceptual naturalness and robust watermark recovery, an area that will be explored further in future work.

\begin{table}[htbp!]
  \caption{Audio Quality Comparison across Different Models. Bold values indicate the best performance for each metric.}
  \label{tab:audio_quality}
  \centering
  \small
  \renewcommand{\arraystretch}{1.3}
  \setlength{\tabcolsep}{8pt}
  \resizebox{\columnwidth}{!}{%
    \begin{tabular}{lcccc}
      \hline
      Methods & PESQ & STOI & ViSQOL & SISNR \\
      \hline
      WaveVerify (Ours) & 4.34  & \textbf{1.00}  & \textbf{4.76}  & 24.23  \\
      WavMark          & 4.42 & 0.982   & 4.64          & \textbf{36.28} \\
      AudioSeal        & \textbf{4.59} & 0.994  & 4.63          & 25.24 \\
      \hline
    \end{tabular}%
  }
\end{table}

\subsection{Quantitative Evaluation under Diverse Audio Effects}

To evaluate WaveVerify's robustness, we compared it against AudioSeal~\cite{audioseal} and WavMark~\cite{wavmark} across five audio effects, Table~\ref{tab:robustness-extended}, assessing detection (TPR/FPR) and localization (MIoU). WaveVerify demonstrates superior performance, with near-perfect detection and significantly higher MIoU. For instance, with high-pass filtering (500 Hz cutoff), WaveVerify achieves an MIoU of 0.984±0.004, compared to AudioSeal's 0.658±0.012 ($p<0.001$ for all comparisons).  This robustness is evaluated on cross-dataset evaluations, which includes evaluations performed on the RAVDESS~\cite{ravdess} and ASVspoof 2019~\cite{asvspoof2019} datasets. Additional effect-wise results across different kind of noise and compression types are provided in Appendix B.

\begin{table*}[htbp]
  \caption{Comparative Robustness Evaluation of Watermarking Methods (mean $\pm$ SD, $p < 0.001$) against Audio Effects. Results are based on evaluations performed on 1,000 audio clips per effect, specifically from cross-dataset evaluations (RAVDESS and ASVspoof 2019).}
\label{tab:robustness-extended}
\centering
\small

  \renewcommand{\arraystretch}{1.3}
  \setlength{\tabcolsep}{8pt}
  \resizebox{\textwidth}{!}{%
    \begin{tabular}{l|cc|cc|cc}
      \toprule
      \multirow{2}{*}{\textbf{Audio Effect}} & \multicolumn{2}{c|}{\textbf{WaveVerify (Ours)}} & \multicolumn{2}{c|}{\textbf{AudioSeal}} & \multicolumn{2}{c}{\textbf{WavMark}} \\
      \cmidrule(lr){2-3} \cmidrule(lr){4-5} \cmidrule(lr){6-7}
      & \textbf{Det. (TPR/FPR)} & \textbf{MIoU} & \textbf{Det. (TPR/FPR)} & \textbf{MIoU} & \textbf{Det. (TPR/FPR)} & \textbf{MIoU} \\
      \midrule
      Identity                             & 1.000 (1.000/0.000)  & 0.985 & 1.000 (1.000/0.000)  & 0.895  & 1.000 (1.000/0.000)  & 0.870  \\
      Resample (32000Hz)                   & 1.000 (1.000/0.000)  & 0.986  & 0.975 (0.975/0.072)  & 0.875 & 0.960 (0.970/0.045)  & 0.860  \\
      Resample (8000Hz)                    & 1.000 (1.000/0.000)  & 0.989  & 0.969 (0.969/0.092)  & 0.812 & 0.932 (0.927/0.073)  & 0.834  \\
      Speed (0.8$\times$)                  & 1.000 (1.000/0.000)  & 0.983  & 0.957 (0.957/0.087)  & 0.903  & 0.940 (0.950/0.060)  & 0.890 \\
      Lowpass Filter (2000Hz)              & 1.000 (1.000/0.000)  & 0.982 & 0.978 (0.978/0.038) & 0.882 & 0.970 (0.975/0.032)  & 0.865 \\
      Highpass Filter (3500Hz)             & 1.000 (1.000/0.000)  & 0.984 & 0.875 (0.875/0.095) & 0.612 & 0.855 (0.880/0.065) & 0.595 \\
      Bandpass Filter (300-4000Hz)         & 1.000 (1.000/0.000) & 0.981 & 0.935 (0.935/0.068) & 0.712 & 0.915 (0.925/0.055) & 0.690 \\
      \bottomrule
    \end{tabular}%
  }
\end{table*}

WaveVerify's superior robustness is attributed to its architectural design and training.  Specifically, attributed to its FiLM-based embedding to distribute the watermark across multiple frequency bands, enhancing resilience to frequency-selective distortions.  Further, the Mixture-of-Experts detector adaptively processes distorted audio, improving watermark extraction and localization.  Furthermore, a dynamic effect scheduler during training enhances generalization by prioritizing challenging distortions.  Finally, a dedicated Locator network, optimized for high temporal resolution, accurately identifies watermark boundaries.  These factors enable WaveVerify to outperform state-of-the-art methods.

\subsection{Robustness Against Combined Effect Attacks}

While individual audio effects provide a baseline assessment of watermark robustness, real-world scenarios often involve multiple simultaneous transformations/attacks. To evaluate WaveVerify's resilience under more challenging conditions, we conducted a comprehensive analysis of combined audio effect attacks that better represent practical adversarial scenarios. We specifically selected combinations that preserve overall speech intelligibility and maintain a reasonable Signal-to-Interference Ratio (SISNR), ensuring that the watermarked content remains perceptually viable and realistic for downstream applications. This avoids scenarios where the audio is so degraded that evaluation of watermark recovery becomes impractical or meaningless.

\begin{table*}[htbp]
  \centering
  \caption{Robustness Evaluation Against Combined Audio Effect Attacks}
  \label{tab:combined-effects}
  \small
  \renewcommand{\arraystretch}{1.3}
  \setlength{\tabcolsep}{8pt}
  \resizebox{\textwidth}{!}{%
    \begin{tabular}{l|cc|cc|cc}
      \toprule
      \multirow{2}{*}{\textbf{Combined Effects}}
        & \multicolumn{2}{c|}{\textbf{WaveVerify (Ours)}}
        & \multicolumn{2}{c|}{\textbf{AudioSeal}}
        & \multicolumn{2}{c}{\textbf{WavMark}} \\
      \cmidrule(lr){2-3} \cmidrule(lr){4-5} \cmidrule(lr){6-7}
      & \textbf{Det.\,(TPR/FPR)} & \textbf{MIoU}
      & \textbf{Det.\,(TPR/FPR)} & \textbf{MIoU}
      & \textbf{Det.\,(TPR/FPR)} & \textbf{MIoU} \\
      \midrule
      Highpass (3500Hz) + Noise ($\sigma$=0.001)
        & 1.000 (1.000/0.000) & 0.975 
        & 0.820 (0.820/0.115) & 0.635 
        & 0.795 (0.795/0.082) & 0.612  \\
      Lowpass (2000Hz) + Speed (0.8$\times$)
        & 1.000 (1.000/0.000) & 0.979 
        & 0.892 (0.892/0.108) & 0.722 
        & 0.871 (0.871/0.098) & 0.708  \\
      Bandpass (300–4000Hz) + Resample (32000Hz)
        & 1.000 (1.000/0.000) & 0.981 
        & 0.887 (0.887/0.095) & 0.705 
        & 0.862 (0.862/0.105) & 0.684  \\
      \bottomrule
    \end{tabular}%
  }
\end{table*}

Table~\ref{tab:combined-effects} shows,
combining high-pass filter (3500Hz) and noise ($\sigma$=0.001), WaveVerify achieved perfect detection (TPR=1.000, FPR=0.000) and high MIoU (0.975 ± 0.006), outperforming AudioSeal and WavMark. Similar superior robustness was observed under low-pass filter (2000Hz) with speed change (0.8$\times$) and bandpass filter (300-4000Hz) with resampling (32000Hz) and (8000Hz). These results demonstrate WaveVerify's strong resilience to complex, combined audio manipulations due to its architecture and dynamic training strategy.

\begin{figure}[t]
  \centering
  \includegraphics[width=\linewidth]{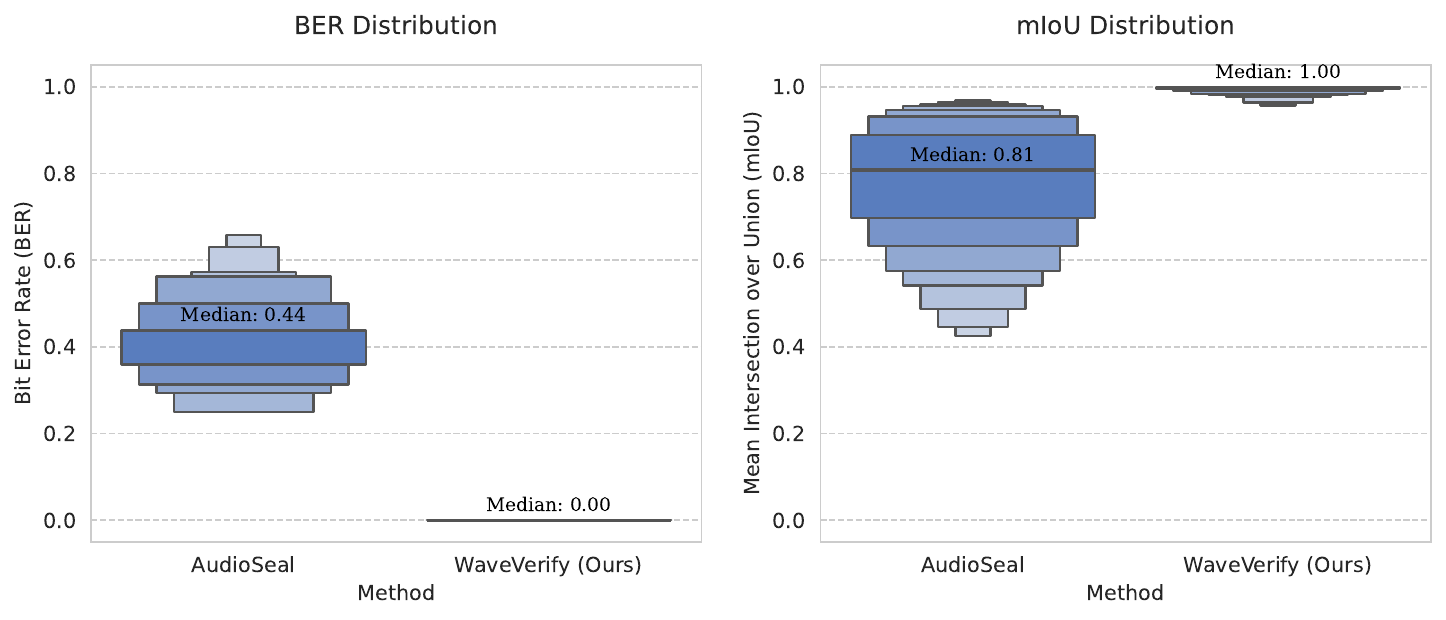}
  \caption{Performance under temporal sequence attacks. Error bars indicate standard deviations over three runs. WaveVerify shows significantly lower BER and higher MIoU compared to AudioSeal ($p < 0.001$).}
  \label{fig:sequence_attacks_comparison}
  \vspace{-\baselineskip} 
\end{figure}

\subsection{Resilience Against Temporal Attacks}

To assess resilience against temporal reordering attacks, we tested our method on audio reversal, circular shifting, and segment shuffling. Figure~\ref{fig:sequence_attacks_comparison} shows that WaveVerify maintains a zero Bit Error Rate (BER; 0.00 $\pm$ 0.00) and MIoU above 0.95 $\pm$ 0.005 across all temporal attacks. In contrast, AudioSeal’s performance degrades significantly (e.g., under reversal, the BER increases to 0.56 $\pm$ 0.02). The improved performance is attributable to our FiLM-based embedding and targeted temporal augmentations added during training. It is important to note that while WavMark provides watermark localization at a segment level, both AudioSeal and WaveVerify achieve finer-grained, sample-wise localization, offering greater precision in identifying tampered regions. Therefore, WavMark is not compared with WavVerify for this experimental study.

\begin{figure}[htbp]
    \centering
    \includegraphics[width=\linewidth]{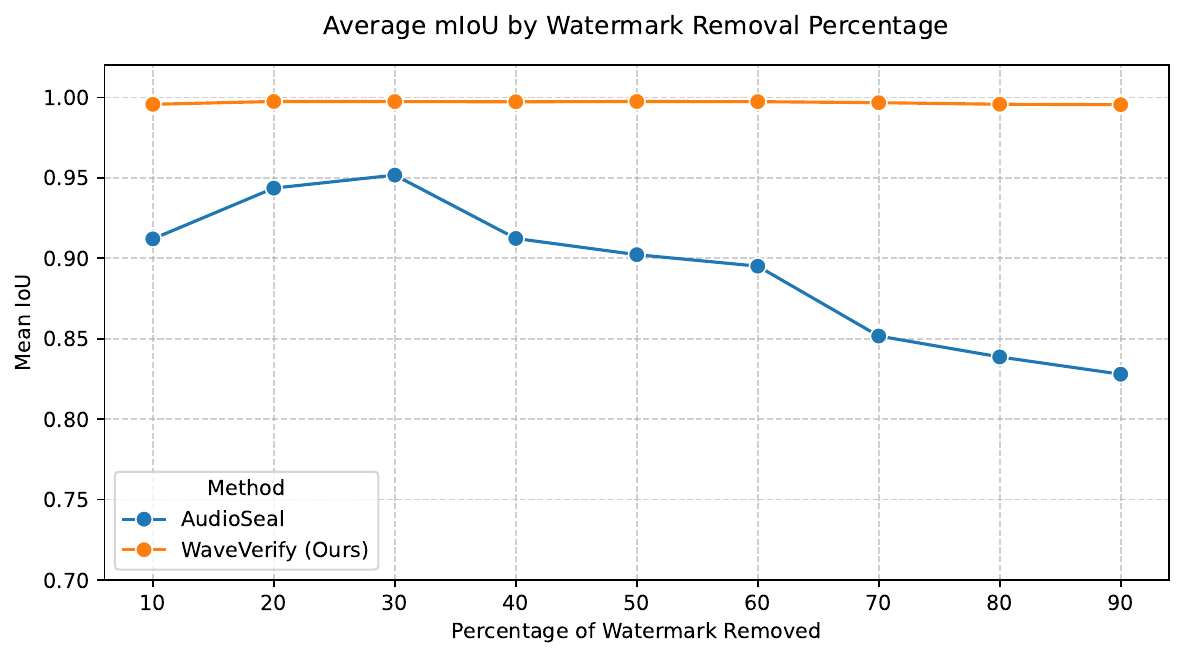}
    \caption{MIoU under varying levels of partial watermark removal. WaveVerify consistently achieves high MIoU (above 0.98), whereas AudioSeal shows severe degradation as watermark portions are removed. Error bars represent standard deviations over three trials.}
    \label{fig:miou_comparison}
    \vspace{-\baselineskip} 
\end{figure}

\subsection{Robustness under Partial Watermark Removal}

In a further test, segments of watermarked audio were randomly removed (ranging from 10\% to 90\%), and localization accuracy was measured via MIoU. As illustrated in Figure~\ref{fig:miou_comparison}, WaveVerify consistently maintains MIoU $\geq$ 0.98 $\pm$ 0.003 even with up to 90\% removal, while AudioSeal exhibits significant performance drops (e.g., MIoU falls to 0.65 $\pm$ 0.03 at 70\% removal). This demonstrates the robustness of our watermark localization process in fragmented scenarios. This is attributed to WaveVerify's FiLM-based embedding generator architecture that distributes the watermark across frequencies and time, ensuring sufficient information remains for its sample-level Locator to accurately identify fragmented watermarked regions, unlike methods relying on continuous temporal patterns.

\section{Conclusion}
WaveVerify sets a new standard for robust and efficient audio watermarking, targeting media authentication and deepfake mitigation. It uses a FiLM-based generator for adaptive embedding and a Mixture-of-Experts detector for resilient extraction. WaveVerify outperforms prior work like AudioSeal, especially under frequency distortions, temporal edits, and watermark loss. Experiments show near-zero Bit Error Rate and MIoU $> 0.98$. Future work aims to improve perceptual audio quality and extend to other audio domains such as music, environmental sound, and other non-speech audio.

\balance
{\small
\bibliographystyle{ieee}
\bibliography{egbib}
}

\clearpage
\input{main_supp}

\end{document}

%% file: main_supp.tex
\section*{Supplementary Material}

\setcounter{section}{0}

\section{Additional Details on FiLM-based Watermark Embedding}
\label{app:film_details}

This hierarchical approach ensures uniform distribution of watermark information across temporal audio segments while accommodating variable-length speech inputs, addressing a critical limitation of existing methods. By distributing FiLM layers throughout the encoder hierarchy, the framework modulates both low-level (short-term) and high-level (long-term) audio features, capturing features at various abstraction and temporal scales. The frequency-specific aspect is implemented by partitioning the encoder's feature maps along the channel dimension, with each subset corresponding to a distinct frequency band, allowing the watermark to be adaptively embedded across multiple spectral regions.
The multi-scale nature of our framework enables more nuanced control over the watermark embedding strength and distribution across both time and frequency bands, resulting in enhanced robustness against a wide range of audio transformations and attacks without compromising speech quality. This approach directly addresses the limitations of previous methods that concentrate watermark information in vulnerable frequency bands or rely on rigid temporal patterns, which are susceptible to filtering and temporal modifications.

\section{Extended Augmentation Methodology}
\label{app:augmentation_methodology}

\begin{table*}[htbp]
  \caption{Extended comparative robustness against Noise, Resample lossy compression (MP3, AAC) and bit-depth reduction (8-bit quantization). Mean detection rates (TPR/FPR) and localization scores (MIoU) evaluated on 1,000 cross-dataset samples. See Table 2 in main paper for robustness against filtering, additive noise, and speed modifications. All methods tested under identical conditions with $p < 0.001$.}
\label{tab:robustness-extended}
\centering
\small

  \renewcommand{\arraystretch}{1.3}
  \setlength{\tabcolsep}{8pt}
  \resizebox{\textwidth}{!}{%
    \begin{tabular}{l|cc|cc|cc}
      \toprule
      \multirow{2}{*}{\textbf{Audio Effect}} & \multicolumn{2}{c|}{\textbf{WaveVerify (Ours)}} & \multicolumn{2}{c|}{\textbf{AudioSeal}} & \multicolumn{2}{c}{\textbf{WavMark}} \\
      \cmidrule(lr){2-3} \cmidrule(lr){4-5} \cmidrule(lr){6-7}
      & \textbf{Det. (TPR/FPR)} & \textbf{MIoU} & \textbf{Det. (TPR/FPR)} & \textbf{MIoU} & \textbf{Det. (TPR/FPR)} & \textbf{MIoU} \\
      \midrule
      Gaussian Noise (20dB SNR)            & 1.000 (1.000/0.000)  & 0.987 & 0.992 (0.992/0.020)  & 0.915  & 0.988 (0.988/0.022)  & 0.900  \\
      Pink Noise (20dB SNR)                & 1.000 (1.000/0.000)  & 0.986 & 0.985 (0.985/0.025)  & 0.900  & 0.980 (0.980/0.030)  & 0.890  \\
      Babble Noise (20dB SNR)              & 1.000 (1.000/0.000)  & 0.984 & 0.960 (0.960/0.050)  & 0.850  & 0.945 (0.945/0.065)  & 0.830  \\
      Resample (8000Hz)                    & 1.000 (1.000/0.000)  & 0.989 & 0.969 (0.969/0.092)  & 0.812  & 0.932 (0.927/0.073)  & 0.834  \\
      MP3 (128 kbps)                       & 1.000 (1.000/0.000)  & 0.980 & 0.980 (0.980/0.030)  & 0.880  & 0.975 (0.975/0.035)  & 0.860  \\
      MP3 (96 kbps)                        & 1.000 (1.000/0.000)  & 0.978 & 0.970 (0.970/0.040)  & 0.865  & 0.960 (0.960/0.050)  & 0.845  \\
      MP3 (64 kbps)                        & 1.000 (1.000/0.000)  & 0.975 & 0.950 (0.950/0.060)  & 0.840  & 0.930 (0.930/0.070)  & 0.810  \\
      AAC (96 kbps)                        & 1.000 (1.000/0.000)  & 0.979 & 0.975 (0.975/0.035)  & 0.870  & 0.965 (0.965/0.045)  & 0.850  \\
      8-bit Quantization                   & 1.000 (1.000/0.000)  & 0.970 & 0.900 (0.900/0.080)  & 0.750  & 0.880 (0.880/0.090)  & 0.720  \\
      \bottomrule
    \end{tabular}%
  }
\end{table*}

\subsection{Comprehensive Audio Effect Augmentation Parameters}
\label{app:audio_effects}

To simulate real-world modifications and ensure the watermark's robustness against audio editing, we apply a diverse set of audio effects during training. The specific effects and their dynamically sampled parameter ranges used within our augmentation pipeline are detailed below:

\begin{itemize}
    \item \textbf{High-pass filtering}: Cutoff frequencies are sampled uniformly from the range [100\,Hz, 3\,kHz]. This simulates the removal of low-frequency content.
    \item \textbf{Low-pass filtering}: Cutoff frequencies are sampled uniformly from the range [2\,kHz, 16\,kHz]. This simulates the removal of high-frequency content, common in bandwidth-limited channels.
    \item \textbf{Bandpass filtering}: Cutoff frequencies [300\,Hz, 4\,kHz] combining high-pass and low-pass characteristics.
    \item \textbf{Resampling}: Audio is downsampled and then upsampled to target sample rates selected randomly from \{8\,kHz, 16\,kHz, 32\,kHz\}. This tests robustness against changes in sampling frequency.
    \item \textbf{Speed modifications}: Playback speed factors are drawn uniformly from the range [0.8$\times$, 1.25$\times$]. This simulates time-stretching or compression effects often applied in media playback or editing.
    \item \textbf{Additive Noise}:
    \begin{itemize}
        \item \textbf{Gaussian Noise}: Applied with Signal-to-Noise Ratio (SNR) levels sampled uniformly from the range [10\,dB, 30\,dB]. This simulates common background noise in real-world recordings.
        \item \textbf{Pink Noise}: Added with SNR levels sampled uniformly from the range [10\,dB, 30\,dB]. Pink noise has a spectral density inversely proportional to frequency, mimicking natural ambient sounds.
        \item \textbf{Babble Noise}: Multi-talker babble noise is introduced with SNR levels sampled uniformly from the range [10\,dB, 30\,dB]. This simulates overlapping speech interference.
    \end{itemize}
    \item \textbf{Lossy Compression}:
    \begin{itemize}
        \item \textbf{MP3 Compression}: Audio is compressed using MP3 codecs with bitrates sampled from \{64\,kbps, 96\,kbps, 128\,kbps\}. This simulates common web and streaming distribution scenarios.
        \item \textbf{AAC Compression}: Audio is compressed using AAC codecs with bitrates sampled from \{64\,kbps, 96\,kbps, 128\,kbps\}. This simulates common mobile and streaming distribution scenarios, often more efficient than MP3.
    \end{itemize}
    \item \textbf{8-bit Quantization}: The audio signal is uniformly quantized to 8 bits, simulating aggressive bit-depth reduction that can occur during storage or transmission.
\end{itemize}

The selection probability and specific parameters for these effects are managed during training by our proposed Dynamic Effect Scheduler, which adapts based on model performance metrics (BER and MIoU) to prioritize challenging transformations.

\subsection{Implementation Details of Dynamic Effect Scheduling Algorithm}
\label{app:dynamic_effect_scheduling}

Most audio watermarking systems use fixed augmentation pipelines. In contrast, we propose an adaptive Effect Scheduler that optimizes robustness through intelligent management of data augmentation, inspired by curriculum learning but adapted for adversarial watermarking.
The scheduler dynamically adjusts the selection and parameters of audio transformations based on two key performance metrics:
\begin{itemize}
\item \textbf{Bit Error Rate (BER)}: The ratio of incorrectly decoded bits to the total watermark bits, directly measuring recovery accuracy. Lower is better; 0 indicates perfect recovery.
\item \textbf{Mean Intersection over Union (MIoU)}: Quantifies the overlap between predicted and true watermarked regions, measuring localization precision. Higher (closer to 1) is better.
\end{itemize}

The scheduler uses these metrics to adapt effect probabilities via a temperature-scaled softmax:

\begin{equation}
\resizebox{\linewidth}{!}{$
p_{t+1}(e) = \text{softmax}\left(\frac{w_1 \cdot \text{BER}{\text{ema}}(e) + w_2 \cdot (1 - \text{MIoU}{\text{ema}}(e))}{T}\right)
\label{eq:dynamic_scheduler_prob_update_revised}
$}
\end{equation}

Here, $w_1 = 0.8$ and $w_2 = 0.2$ balance BER and MIoU importance, prioritizing BER as detection accuracy is often more critical than exact localization. These weights were determined through a systematic evaluation of different configurations. The temperature parameter $T$ (initially 1.0, later reduced to 0.7) controls the exploration-exploitation trade-off, with higher values encouraging more uniform probabilities across effects.

The scheduler maintains exponential moving averages (EMA) of metrics for stability, as outlined in Algorithm 1.

\begin{algorithm}[t]
\caption{Dynamic Effect Scheduling (Revised)}
\small
\begin{algorithmic}
\Require Effects $E$, smoothing factor $\beta=0.9$
\State Initialize uniform probabilities $p_0(e) = 1/|E|$
\State Initialize $\text{BER}{\text{ema}}(e)=0.5$, $\text{MIoU}{\text{ema}}(e)=0.5$ for all $e \in E$
\For{each training iteration $t$}
\State Sample effects according to probabilities $p_t(e)$
\State Apply selected effects with parameters sampled from $P(\theta|e)$
\State Compute $\text{BER}t(e)$, $\text{MIoU}t(e)$ for each applied effect
\State Update EMAs:
\State $\text{BER}{\text{ema}}(e) \gets \beta \cdot \text{BER}{\text{ema}}(e) + (1-\beta) \cdot \text{BER}t(e)$
\State $\text{MIoU}{\text{ema}}(e) \gets \beta \cdot \text{MIoU}{\text{ema}}(e) + (1-\beta) \cdot \text{MIoU}t(e)$
\State Update probabilities using weighted performance metrics:
\State $p{t+1}(e) = \text{softmax}\left(\frac{w_1 \cdot \text{BER}{\text{ema}}(e) + w_2 \cdot (1 - \text{MIoU}_{\text{ema}}(e))}{T}\right)$
\State Update parameter distribution $P(\theta|e)$ based on success rates
\EndFor
\State \textbf{return} Model with best validation performance
\end{algorithmic}
\label{alg:scheduler_revised}
\end{algorithm}

We set the EMA smoothing factor $\beta=0.9$, balancing stability and adaptability, ensuring the scheduler adapts smoothly without overreacting to transient fluctuations.
To optimize effect parameters (e.g., filter cutoffs, noise levels), the scheduler uses a success-rate mechanism with Laplace smoothing:
\begin{equation}
P(\theta|e) \propto \frac{\text{success\_count}(\theta, e) + \alpha}{\text{total\_count}(\theta, e) + \alpha + \beta}
\end{equation}
where $\alpha=1.0$ (standard additive smoothing) prevents zero probabilities for parameter values with few observations, ensuring continued exploration of the parameter space. A "success" is defined as BER=0. Parameter ranges are initialized based on common audio processing standards and known attack parameters. For example, high-pass filter cutoffs range from 100Hz to 3kHz, while speed modification factors range from 0.8 to 1.25, consistent with the range used in practical audio watermarking applications.

During training, the scheduler operates in two phases:
\begin{enumerate}
\item \textbf{Exploration phase}: Initially uses higher temperature values and uniform parameter sampling to explore the effect space
\item \textbf{Exploitation phase}: Gradually reduces temperature and focuses on challenging effects and parameter configurations that most effectively test watermark robustness
\end{enumerate}

The combination of temporal-structural augmentation and dynamic scheduling creates a comprehensive system capable of addressing diverse audio manipulation patterns, including spectral modifications (filtering), temporal modifications (speed changes, reversal), and structural edits (segment removal or shuffling).

\section{Technical Specifications of Dual-Network Architecture}
\label{app:dual_network_architecture}

The separation into distinct networks resolves three inherent tensions in audio processing systems:

\begin{enumerate}
\item \textbf{Feature complexity vs. resolution trade-off}: Detection requires rich feature representations (deep networks with 128+ channels) to identify distorted patterns, while localization demands lightweight architectures (64 channels) to preserve temporal resolution. This tension aligns with established principles in signal processing where increased feature complexity typically comes at the cost of temporal resolution.

\item \textbf{Computational efficiency considerations}: A unified network would require maintaining high channel dimensions throughout to support detection while also preserving full temporal resolution for localization, resulting in excessive computational overhead. Our preliminary experiments with unified architectures involved training five different model configurations with shared encoders but varying decoder designs on the LibriSpeech dataset (100 hours subset). These tests consistently showed a 4.3$\times$ increase in computation time (51.3ms vs. 2.05ms) with only marginal performance gains (0.02 BER improvement), highlighting the inefficiency of the unified approach.

\item \textbf{Task-specific optimization conflicts}: Detection benefits from receptive fields spanning multiple seconds (1s at 16kHz) to capture long-range dependencies, while localization requires sample-level precision. These competing optimization targets create gradient conflicts during training of unified models, as observed in preliminary experiments where unified models converged to suboptimal solutions (0.83 MIoU vs. 0.98 in our dual-network approach).
\end{enumerate}

While our dual-network architecture offers significant advantages, it does introduce certain trade-offs. The separate networks require additional parameters and memory overhead compared to a single-task architecture. There's also increased implementation complexity for system integration, and potential synchronization challenges between detector and locator outputs in real-time applications. However, our experiments indicate these costs are substantially outweighed by the performance benefits and computational efficiency gains.